\documentclass{PoS}

\usepackage{color}

\newcommand{\be}{\begin{equation}}
\newcommand{\ee}{\end{equation}}
\newcommand{\bea}{\begin{eqnarray}}
\newcommand{\eea}{\end{eqnarray}}

\newcommand{\MKK}{M_{\rm KK}}

\newcommand{\Tr}{{\rm Tr}\,}

\def\PDG{\cite{PDG15}}
\def\BPR{\cite{Brunner:2015oqa}}
\def\BR{\cite{1504.05815}}
\def\BReep{\cite{1510.07605}}

\def\WSS{Witten-Sakai-Sugimoto}

\definecolor{Red}{rgb}{0.7,0,0}
\definecolor{Green}{rgb}{0,0.4,0}
\definecolor{Blue}{rgb}{0,0,0.7}
\definecolor{gray}{rgb}{0.5,0.5,0.5}

\def\red{\color{red}}

\def\gray{\color{gray}}

\title{Glueball decay patterns\\ in top-down holographic QCD}

\ShortTitle{Glueball decay patterns in top-down holographic QCD}

\author{Frederic Br\"unner, Denis Parganlija, and \speaker{Anton Rebhan}
\\
       Institute f\"ur Theoretische Physik, Technische Universit\"at Wien,\\ Wiedner Hauptstra{\ss}e 8-10, A-1040 Vienna, Austria\\
        E-mail: \email{bruenner,denisp,rebhana@hep.itp.tuwien.ac.at}}

\abstract{We discuss our results on scalar glueball decay in the top-down holographic \WSS\ model
for low-energy QCD and compare with available experimental data, which appear to
disfavor the glueball candidate $f_0(1500)$ but seem to be perfectly consistent with
interpreting $f_0(1710)$ as a nearly unmixed glueball. The holographic model moreover makes
definite predictions for future experiments.}

\FullConference{The European Physical Society Conference on High Energy Physics\\
		22--29 July 2015\\
		Vienna, Austria}

\begin{document}

\section{Introduction}

Quantum chromodynamics (QCD) predicts glueballs,
color-neutral bound states of gluons, to show up as 
flavor singlet mesons beyond those provided
by the quark model \cite{Fritzsch:1972jv},
but their experimental status is still very unclear \cite{Klempt:2007cp
}.
The mass of the lowest glueball state
with $J^{PC}=0^{++}$ is estimated by lattice gauge theory \cite{Morningstar:1999rf,Gregory:2012hu}
to be in the range 1.5-1.8 GeV, in which the Particle Data Group \PDG\ lists two fairly narrow 
isoscalar $0^{++}$ mesons as established,
$f_0(1500)$ and $f_0(1710)$. Together with the wider resonance $f_0(1370)$ these are often described 
in phenomenological models as
mixtures of the two isoscalar $q\bar q$ states $u\bar u+d\bar d$ and $s\bar s$ and the lightest scalar glueball
\cite{Amsler:1995td,Lee:1999kv,Close:2001ga,Amsler:2004ps,Close:2005vf,Giacosa:2005zt,Albaladejo:2008qa,Mathieu:2008me,Janowski:2011gt,Janowski:2014ppa,Cheng:2015iaa,Close:2015rza,Frere:2015xxa}, but there is not yet agreement whether the glueball
is the dominant component of $f_0(1500)$ or $f_0(1710)$. 

Clearly, further input from a first-principles approach such as lattice QCD with dynamical quarks would be needed, which
is however computationally difficult since glueball operators are intrinsically noisy. Moreover, decays are Minkowski-space phenomena, while lattice gauge theory needs to be set up in Euclidean space.


Another first-principles approach to strongly interacting nonabelian gauge theories is provided (for certain
theories in certain limits) by gauge/gravity duality, whenever an explicit string-theoretic (top-down) construction is
available. Such a construction is in fact to some degree
available for the low-energy limit of nonsupersymmetric large $N_c$ Yang-Mills-theories through
Witten's model \cite{Witten:1998zw} based on $N_c$ D4 branes in type-IIA supergravity. There the dual gauge theory is
nonconformal 5-dimensional super-Yang-Mills theory, where supersymmetry is broken completely through
a Kaluza-Klein compactification on a circle with radius $\MKK^{-1}$. At scales much smaller than $\MKK$ one
obtains 4-dimensional pure-glue large-$N_c$ Yang-Mills theory, whose glueball spectrum has been worked out in
\cite{Constable:1999gb
}. This model has been extended to include $N_f\ll N_c$ chiral quarks
in a D8 brane construction by Sakai and Sugimoto \cite{Sakai:2004cn
} 
which features
chiral symmetry breaking $\mathrm{U}(N_f)_L\times\mathrm{U}(N_f)_R
\to\mathrm{U}(N_f)_V$, accompanied by massless Nambu-Goldstone bosons, with the one corresponding to
the anomalous U(1)$_A$ 
receiving a mass through the Witten-Veneziano mechanism.

The \WSS\ (WSS) model has been remarkably successful in reproducing qualitatively, as well as to a surprising extent
quantitatively, many aspects of low-energy QCD, while involving essentially 
only one dimensionless parameter, the 't Hooft coupling $\lambda$
at the scale $\MKK$, where validity of the supergravity approximation requires $\lambda\gg1$.
Extrapolating down to finite $N_c=3$, 
and fixing $\MKK$ by the $\rho$ mass,
one obtains for example
for the decay rates of $\rho$ and $\omega$ mesons
\be
\Gamma(\rho\to2\pi)/ m_\rho = 0.1535\dots 0.2034,\quad
\Gamma(\omega\to3\pi)/ m_\omega = 0.0033\ldots 0.0102,
\ee
when $\lambda$ is varied between 16.63 and 12.55 such that either the pion decay constant \cite{Sakai:2004cn} 
or the string tension in large-$N_c$ lattice simulations is matched \cite{Rebhan:2014rxa}.
Encouragingly, this covers the experimental values, 0.191(1) and
0.0097(1), respectively. 

It thus seems interesting to explore the predictions of this model for glueball decays.
Since the WSS model is formulated in the 't Hooft limit, it predicts narrow decay widths $\propto \lambda^{-1}N_c^{-2}$
and also $1/N_c$ suppressed mixing of glueballs
with $q\bar q$ states \cite{Hashimoto:2007ze}.
By contrast, holographic (bottom-up) models in the Veneziano limit of QCD
($N_f/N_c$ is kept fixed as $N_c\to\infty$) \cite{Arean:2013tja} naturally have large mixing \cite{Iatrakis:2015rga};
in the WSS model we shall ignore mixing which is absent at our level of approximation.

\section{Glueball decay in the \WSS\ model}

In \BPR\ the effective action of the scalar and tensor glueball modes of the WSS model
has been worked out in detail, correcting and generalizing previous work \cite{Hashimoto:2007ze}.
The scalar $0^{++}$ glueball has vertices with $q\bar q$ mesons of the form (ignoring derivatives and dropping Lorentz indices)
\be\label{GverticesLO}
G\Tr(\pi\pi), \quad G\Tr(\rho\rho), \quad G\Tr(\rho[\pi,\pi]), \quad G\Tr([\pi,\rho]^2),
\quad G\Tr(\rho[\rho,\rho]),\quad G\Tr([\rho,\rho]^2),
\ee
where $\pi$ and $\rho$ represent the pseudoscalar and vector meson nonets, but for a lowest-lying glueball
with mass well below 2 GeV only $\rho$ and $\omega$ will be relevant. 

The Witten model has in fact more scalar glueball modes than suggested by lattice results. 
One tower of modes, which contains the lightest mode with $\approx 0.901\MKK\approx 855$~MeV, is associated with an
``exotic'' \cite{Constable:1999gb} graviton
polarization involving the direction of Kaluza-Klein compactification. In \BPR\ we have argued that this
mode ($G_E$) should be discarded from the spectrum, since the next-lightest scalar, which is a predominantly dilatonic
mode ($G_D$)
is more closely related to the standard glueball operator of QCD. Moreover, the latter has a smaller decay
width, which seems unnatural when it should figure as the first excited scalar state.\footnote{%
Another, more speculative possibility would be that the exotic scalar mode
corresponds to a broad glueball component of the $\sigma$-meson in line
with the scenario of Ref.~\cite{Narison:1996fm
},
which features a broad glueball around 1 GeV and a narrower one around 1.5 GeV.}
The lightest $G_D$ mode turns out to have a mass of $\approx 1.567\MKK\approx 1487$~MeV,
which is in the ballpark indicated by lattice QCD, very close to the mass of $f_0(1500)$,
but also just 14\% lighter than $f_0(1710)$.

In \BPR\ we have calculated the decay pattern of the $G_D$ mode by using the vertices
(\ref{GverticesLO}) in tree-level amplitudes with the mass of the scalar glueball
set alternatively to that of $f_0(1500)$ and $f_0(1710)$ in order to see which, if any, would
be compatible with a nearly pure glueball interpretation.\footnote{The extrapolation
of the mass of the glueball was accompanied by a rescaling of the dimensionfull
glueball couplings such that the dimensionless ratio
$\Gamma(G_D\to\pi\pi)/M$ is unchanged for massless $\pi$.
Decays involving the massive vector mesons do however change significantly by phase space factors, because $f_0(1710)$ is above the $2\rho$ threshold.}
The results are shown in Table \ref{tabrates}. 

In the case of $f_0(1500)$, the
decay rate into two pions is underestimated my more than a factor of 2, while
the dominant decay mode of $f_0(1500)$ into four pions comes out as about an order of
magnitude too small.\footnote{Because $G_D$ at the mass of $f_0(1500)$
is below the 2$\rho$ threshold, it may be that the evaluation at its nominal mass underestimates
the decay to four pions. Indeed, performing an average over a mass distribution 
from ${M-3\Gamma_{\rm tot}}$ to ${M+3\Gamma_{\rm tot}}$ (that is about 1.2-1.8 GeV) with a
simple Breit-Wigner distribution 
\[
\Gamma_i^{\rm BW}=\frac1{\mathcal N} \int_{M-3\Gamma_{\rm tot}}^{M+3\Gamma_{\rm tot}} dx \frac{M^2 \Gamma_{\rm tot}}{(x^2-M^2)^2+M^2 \Gamma_{\rm tot}^2}\Gamma_i(x),\quad
\mathcal N=\int_{M-3\Gamma_{\rm tot}}^{M+3\Gamma_{\rm tot}} dx \frac{M^2 \Gamma_{\rm tot}}{(x^2-M^2)^2+M^2 \Gamma_{\rm tot}^2},
\]
the partial width into four pions increases 
from $(0.003$-$0.005)M$ to $(0.005$-$0.007)M$, so it remains very small compared to the experimental
value of $f_0(1500)$ (Table \ref{tabrates}). 
For $G_D$ at the mass of $f_0(1710)$, however, the difference
between two procedures is tiny compared to the theoretical uncertainty parametrized by the variation of $\lambda$.\label{fnBW}}
In \BPR\ we have also calculated the decay into
$4\pi^0$, which is parametrically of higher order and therefore strongly suppressed in
the WSS model. The prediction for a pure glueball with the mass of $f_0(1500)$ is 
$\Gamma(4\pi^0)/M=4\times 10^{-6}\ldots 3\times 10^{-5}$. It should therefore be almost unobservably small, but according to
\PDG\ this decay mode is seen. In \cite{Alde:1987ki} one finds $\Gamma(f_0(1500)\to4\pi^0)/\Gamma(f_0(1500)\to2\eta)=0.8(3)$, which would correspond to
a mismatch with the holographic prediction by two orders of magnitude.



\begin{table}
\begin{tabular}{lrrr}
\hline
decay &  $\Gamma/M$ {(exp. \protect\PDG)}  & (WSS massless \protect\cite{Brunner:2015oqa}) & (WSS massive \protect\cite{1504.05815})\\
\hline
$f_0(1500)$ (total) & 0.072(5)  & \red 0.027\ldots0.039
& 0.057\ldots0.079
\\
$f_0(1500)\to4\pi$ & 0.036(3)  &  $\displaystyle (^{**}) \left\{ \red 0.003\ldots 0.005 \atop 0.005\ldots 0.007 \right.$ &  
$\displaystyle\left\{ \red 0.003\ldots 0.005 \atop 0.005\ldots 0.007 \right.$ \\
$f_0(1500)\to2\pi$ & 0.025(2)  & \red 0.009\ldots0.012 & \red 0.010\ldots0.014\\
$f_0(1500)\to 2K$ & 0.006(1)  & \red 0.012\ldots0.016 & \red 0.034\ldots0.045\\
$f_0(1500)\to 2\eta$ & 0.004(1)  & 0.003\ldots0.004 & \red 0.010\ldots0.013\\
$f_0(1500)\to \eta\eta'$ & 0.0014(6)  & \red 0 & $(^{***})$ 0\\
\hline
$f_0(1710)$ (total) & 0.081(5) 
& 0.059\ldots0.076 & 0.083\ldots0.106 \\
$f_0(1710)\to 2K$ &  $(^{*})$ 
\gray 0.029(10) 
& \red 0.012\ldots0.016 & 0.029\ldots0.038 \\[8pt]
$f_0(1710)\to 2\eta$ & \gray 0.014(6) 
& \red 0.003\ldots0.004 & 0.009\ldots0.011 \\[8pt]
$f_0(1710)\to2\pi$ & \gray 0.012($+5\atop-6$) 
& 0.009\ldots0.012 & 0.010\ldots0.013 \\[8pt]
$f_0(1710)\to2\rho,\rho\pi\pi\to4\pi$ & ? & 0.024\ldots 0.030 & 0.024\ldots 0.030 \\ 
$f_0(1710)\to2\omega$ & \gray 0.010($+6\atop-7$) 
& 0.011\ldots 0.014 & 0.011\ldots 0.014 \\ 
$f_0(1710)\to \eta\eta'$ & ?  & 0 & $(^{***})$ 0\\
\hline
\end{tabular}
\caption{Experimental data from Ref.~\PDG\ for the decay rates of the glueball candidates $f_0(1500)$ and
$f_0(1710)$ compared to the results obtained for a pure glueball $G_D$ of same mass in the chiral Witten-Sakai-Sugimoto model \BPR\ 
and for the extrapolation to finite quark masses in \cite{1504.05815}. Red color indicates a significant discrepancy with a pure-glueball interpretation of the respective $f_0$ meson. \newline
$(^{*})$ Gray color indicates that the PDG ratios for various decay rates have been combined
with the branching ratio ${\rm Br}(f_0(1710)\to KK)=0.36(12)$ from Ref.~\cite{Albaladejo:2008qa}.\newline
$(^{**})$ We give two different results for the decay rate $f_0(1500)\to4\pi$ depending on whether the rate
is evaluated as in \BPR\ at the nominal mass of 1504 MeV (upper entry) or averaged by a Breit-Wigner distribution with
experimental width of 109 MeV (cf.~footnote \protect\ref{fnBW}), which increases the rate somewhat since $f_0(1500)$ is not far from the $2\rho$ threshold
(while making only little difference in the case of $f_0(1710)$).\newline
$(^{***})$ If one relaxes the assumption in Ref.~\cite{1504.05815} of a universal coupling of all pseudoscalar mass terms, nonzero $\eta\eta'$ rates are possible in the WSS model with finite quark masses. In the case of $f_0(1710)$ an upper limit of $\Gamma(\eta\eta')/\Gamma(\pi\pi)\lesssim 0.04$ (i.e.\
$\Gamma(\eta\eta')/M \lesssim 0.0005$) is obtained, if one requires that 
prediction for $\Gamma(\pi\pi)/\Gamma(KK)$ remains within the current experimental error bar \cite{1510.07605}. For $f_0(1500)$
one can in fact also find a range of parameters, where the various ratios of decays into two pseudoscalars including $\eta\eta'$ are roughly reproduced. However, the absolute values of the partial decay widths are then all too small by a factor between 2 and 3. With the most important rate into 4 pions being even more strongly underestimated, the total width $\Gamma/M$ is then matched even more poorly than in the massless WSS model. (In this case substantial mixing with $q\bar q$ states would have to account for the decay pattern.)
}
\label{tabrates}
\end{table}

However, $f_0(1500)$ is anyway rarely considered a nearly pure glueball in the phenomenological literature; 
if taken as predominantly glue, it usually has substantial admixture of $q\bar q$.
On the other hand, models which regard $f_0(1710)$ as the preferred glueball candidate 
do so sometimes with rather small $q\bar q$ components \cite{Janowski:2014ppa,Cheng:2015iaa}.
Our results for the decay rate $G_D\to\pi\pi$ indeed look quite compatible with $f_0(1710)$,
if we combine the result quoted by \PDG\ for $\Gamma(\pi\pi)/\Gamma(KK)$ with
the branching ratio ${\rm Br}(f_0(1710)\to KK)=0.36(12)$ from Ref.~\cite{Albaladejo:2008qa}.

The dominant decay mode of $f_0(1710)$, however, is $f_0(1710)\to KK$,
which strongly violates flavor symmetry one would naively expect in glueball decay.
A ratio $\Gamma(\pi\pi)/\Gamma(KK)$ much lower than 3/4 is frequently attributed to ``chiral suppression''
of glueball decay \cite{Carlson:1980kh,Sexton:1995kd,Chanowitz:2005du} according to
which the decay amplitude should be proportional to (current) quark masses. The perturbative
reasoning appears questionable \cite{Frere:2015xxa}, in particular in view of effective quark masses 
produced by chiral symmetry breaking. Indeed, in the massless WSS model glueballs are able
to decay into massless pions. There an enhancement of the decays into heavier pseudoscalars
may arise through additional couplings of glueballs to the mass terms, which are in fact
inevitable when the latter have a diffeomorphism invariant form in the bulk geometry.
In \BR\ two of us have worked out these additional couplings under the assumption of an
additional symmetry which corresponds to the absence of a $G_D\eta\eta'$ coupling.
The corresponding results are given in the rightmost column of Table \ref{tabrates} and
are found to reproduce the experimental ratio $\Gamma(\pi\pi)/\Gamma(KK)$ for $f_0(1710)$
within one sigma. (This was extended recently in \BReep\ where an upper limit on
$\Gamma(\eta\eta')$ was derived from the requirement that the flavor asymmetry
observed in $\Gamma(\pi\pi)/\Gamma(KK)$ remains within the current experimental error.)

If $f_0(1710)$ is indeed to be identified with a nearly unmixed glueball, the WSS model
also predicts a significant branching ratio into four pions as well as into two $\omega$
mesons. The latter decay mode is indeed considered as ``seen'' by the Particle Data Group \PDG,
namely in radiative decays of $J/\psi$ with a ratio $\Gamma(2\omega)/\Gamma(2\pi)=0.78(32)$,
which is perfectly consistent with the holographic result shown in Table \ref{tabrates}.
In contrast to the situation for $f_0(1500)$, the WSS model also predicts a substantial
branching ratio into four pions at the level of $\Gamma(4\pi)/\Gamma(2\pi)\approx 2.4$.
This decay mode is not yet observed in experiment, but is in fact being studied
by the CMS-TOTEM experiment through $f_0(1710)\to2\rho^0$ \cite{Osterberg:2014mta}. 
It will be interesting to see how the holographic prediction of 
$\Gamma(2\pi^+2\pi^-)/\Gamma(2\pi)\approx 0.8$ compares with data.



\begin{acknowledgments}
This work was supported by the Austrian Science
Fund FWF, project no. P26366, and the FWF doctoral program
Particles \& Interactions, project no. W1252.
\end{acknowledgments}

\bibliographystyle{JHEP}
\bibliography{../glueballdecay}

%

\end{document}